\documentclass[final, runningheads]{llncs}
\usepackage[T1]{fontenc}
\usepackage{hyperref}
\usepackage{graphicx}
\newcommand{\colortext}{\color{black}}
\usepackage{amsmath}
\newcommand{\itadata}{\footnotesize \textsl{ITADATA2024: The 3$^{\text{rd}}$ Italian Conference on Big Data and Data Science}}
\usepackage{fancyhdr}

\usepackage{tcolorbox}
\usepackage{xspace}
\usepackage{amsmath}
\tcbuselibrary{skins,breakable}
\usepackage{bold-extra}
\usepackage{colortbl}
\usepackage{xcolor}
\usepackage{fontenc}   
\usepackage{soul}   
\newcommand{\Adult}{\textsc{Adult}\xspace}
\newcommand{\Italia}{\textsc{Italia}\xspace}
\usepackage{multirow}
\usepackage{tikz}

\usetikzlibrary{shadows}

\fancypagestyle{empty}{\fancyhf{}\fancyhead[C]{\itadata}
  \fancyhead[R]{\includegraphics[width=.05\textwidth]{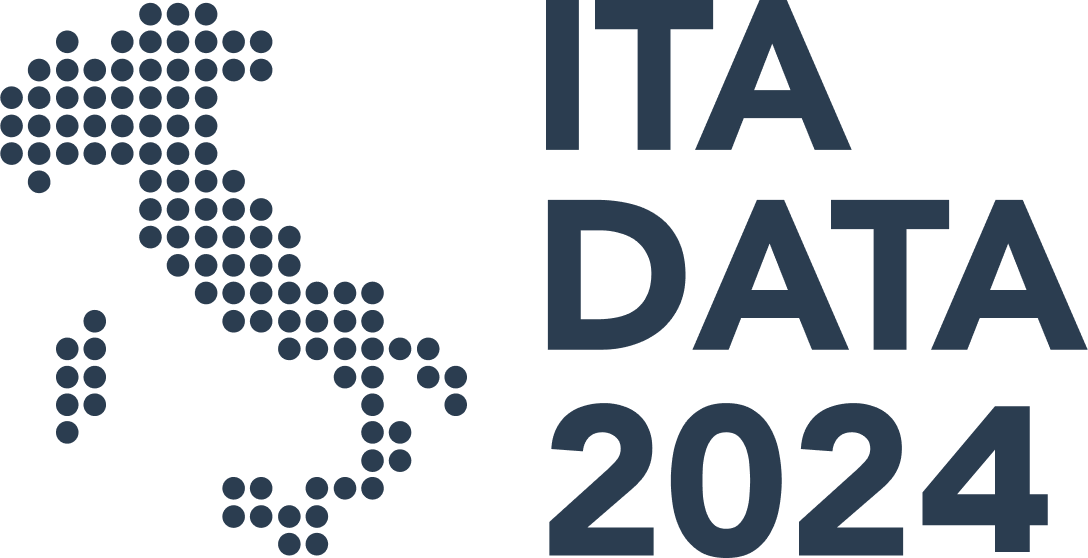}}}

\makeatletter
\begin{document}
\title{Augmenting Anonymized Data with AI: Exploring the Feasibility and Limitations of Large Language Models in Data Enrichment}

\author{Stefano Cirillo\inst{1}\orcidID{0000-0003-0201-2753} \and
Domenico Desiato\inst{2}\orcidID{0000-0002-6327-459X}\and 
Giuseppe Polese\inst{1}\orcidID{0000-0002-8496-2658} \and
Monica Maria Lucia Sebillo\inst{1}\orcidID{0000-0003-3731-6415}\and
Giandomenico Solimando\inst{1}\orcidID{0009-0000-6627-8820}}
\authorrunning{Cirillo et al.}
\institute{Department of Computer Science, University of Salerno \\
\email{\{scirillo,gsolimando,gpolese,msebillo\}@unisa.it}
\and
Department of Computer Science, University of Bari Aldo Moro\\
\email{\{domenico.desiato\}@uniba.it}\vspace{-0.7cm}}
\maketitle              \begin{abstract}
Large Language Models (LLMs) have demonstrated advanced capabilities in both text generation and comprehension, and their application to data archives might facilitate the privatization of sensitive information about the data subjects. In fact, the information contained in data often includes sensitive and personally identifiable details. This data, if not safeguarded, may bring privacy risks in terms of both disclosure and identification. Furthermore, the application of anonymisation techniques, such as k-anonymity, can lead to a significant reduction in the amount of data within data sources, which may reduce the efficacy of predictive processes. In our study, we investigate the capabilities offered by LLMs to enrich anonymized data sources without affecting their anonymity. To this end, we designed new ad-hoc prompt template engineering strategies to perform anonymized Data Augmentation and assess the effectiveness of LLM-based approaches in providing anonymized data. To validate the anonymization guarantees provided by LLMs, we exploited the pyCanon library, designed to assess the values of the parameters associated with the most common privacy-preserving techniques via anonymization. Our experiments conducted on real-world datasets demonstrate that LLMs yield promising results for this goal.

\keywords{Large Language Models \and Data Privacy \and Data Augmentation \and k-Anonimity \and Multimodal Prompt Engineering\vspace{-0.3cm}}
\end{abstract}
\section{Introduction}
The advances in information technology have led to numerous benefits and opportunities for industries, individuals, and society due to the increasing availability of data. The use of analytics approaches has led to the development of increasingly sophisticated applications, spanning from personalized medicine and e-commerce to crowd management and fraud detection. However, these applications have also introduced new privacy and ethical challenges. Data typically holds large amounts of personally identifiable information, e.g., criminal records, shopping habits, credit and medical history, enabling mass surveillance and profiling programs and raising several privacy issues. To prevent these issues from arising, data protection and privacy frameworks usually define strict requirements for the collection and processing of personally identifiable information.

The General Data Protection Regulation (GDPR) has required organizations to collect, process, and share personal data only for legitimate and lawful purposes and to identify privacy risks that can affect data subjects periodically. Employing all measures and procedures to protect personally identifiable information can be costly and may lead to degradation of the quality of the information.
Although anonymisation approaches reduce the risk of re-identification, they can significantly degrade the utility of the data. The resulting loss of data and the introduction of noise mean that predictive models trained on such anonymised datasets often perform poorly in comparison to those trained on raw data. Therefore, it is necessary to define approaches for augmenting data by ensuring their anonymisation.
In this context, large language models (LLMs) offer promising capabilities in generating data for augmenting anonymised data sources. 

These models can generate synthetic datasets maintaining the statistical properties of the original data without compromising individual privacy. By creating high-fidelity synthetic data, LLMs can provide datasets that retain the utility required for effective machine learning while ensuring anonymity.

In this work, we investigate the ability of LLMs to generate and augment anonymized datasets. In particular, we aim to provide insights into the effectiveness of LLM-based approaches to enrich anonymized data 
using new ad-hoc prompting template engineering strategies. To verify the anonymization guarantees offered by LLMs, we adopt a well-known Python library called pyCanon \cite{sainzpardo2022pyCanon}
that assesses the values of the parameters associated with the most common privacy-preserving techniques via anonymization.

The main contributions of the proposed study are summarized as follows: 
\begin{itemize}
\item A new multimodal prompt engineering approach for interacting with LLMs by including real datasets in the prompts;
\item A formalization of the problem of augmenting anonymized data with LLMs;
\item A large evaluation of the most recent LLMs, such as ChatGPT and Claude 3 models, in generating data by preserving the anonymity of the data.
\end{itemize}
The remainder of the paper is organized as follows. Section~\ref{sec:rw} discusses recent related work in the area. Section \ref{sec:Materials_Methods} provides an overview of the datasets, the Large Language Models, and the Anonymization Techniques used in our study. Section \ref{sec:methodology} describes the proposed strategy for augmenting anonymized data using LLMs, whereas Section~\ref{sec:exp} presents experimental results. Finally, Section~\ref{sec:conclusion} provides the conclusion and directions for future work.

\section{Related Work} \label{sec:rw}
Several anonymization techniques have been proposed to facilitate the sharing of sensitive data \cite{ML20}.
One of the most-known techniques is the $k$-anonymity \cite{SS98}, which requires each record in the data to be indistinguishable from at least $k-1$ other records. 
$k$-anonymity is commonly achieved through generalization, which involves replacing attribute values with more general values based on an attribute taxonomy, and suppression, which entails deleting or masking attribute values \cite{ML20}. 
Different approaches utilize various generalization strategies and schemes. 
The process of generalization can be applied to data either globally or locally \cite{Zigomitros2020}. Global schemes involve using the same level of generalization for all attributes, while local schemes allow for different levels of generalization to be applied to each attribute. While ensuring the privacy of individual records in the data, generalization leads to loss of information \cite{EHK2020}. 
While $k$-anonymity protects against identity disclosure, often it does not guarantee sufficient protection against the disclosure of sensitive attributes. As a result, several other anonymization techniques have been defined, such as $\ell$-diversity, $t$-closeness, $m$-confidentiality, and $p$-probabilistic \cite{Zigomitros2020} that take into account the semantic closeness and distribution of the values of sensitive attributes. 
Mondrian has been extended to use value generalization hierarchies \cite{LDR062} and enable $l$-diversity \cite{FKA21}. A data anonymization algorithm ensuring $k$-anonymity and differential privacy was introduced in \cite{BKP18}, using attribute taxonomies and randomized best-first search through generalization hierarchies. The top-down greedy (TDG) anonymization approach \cite{XWP06} iteratively performs binary data partitioning, using a heuristic to split data into equivalence classes and the normalized certainty penalty (NCP) to evaluate information loss.
Access to high-quality data is critical for successful AI model training. These issues make it difficult to achieve optimal results in a variety of AI applications, and can significantly reduce the development and performance of deep learning models. In this context, it is crucial to investigate strategies to perform anonymized Data Augmentation in order to mitigate the low availability of data. In \cite{10.1007/978-3-030-00536-8_1} the authors propose two strategies to generate synthetic medical images with the use of a Generative Adversarial Network (GAN) to train deep learning models. In particular, proposes a method to generate synthetic magnetic resonance imaging (MRI) with brain tumours, to address the imbalance of the available medical imaging datasets in the literature. The results achieved demonstrate the effectiveness of the strategies by improving the performance of the models. In \cite{Bissoto_2021_CVPR} the authors investigate the application of GANs for skin lesion analysis, addressing the challenge of limited training data. They investigate GAN utilization for both data augmentation and anonymization. The results show the effectiveness of the approach in addressing the challenges of in-distribution anonymised sets that can support the availability of limited anonymised data. 
The proposed study explores the effectiveness of Large Language Models (LLMs) in generating anonymized data. Unlike traditional techniques, which often rely on deterministic algorithms and predefined rules to strip personally identifiable information from datasets, LLMs should leverage advanced NLP capabilities to create synthetic data that retains the statistical properties and utility of the original data while ensuring privacy. 

\section{Materials and Methods}\label{sec:Materials_Methods}
Recently, LLMs have been extremely popular thanks to their ability to understand and generate human-like text. In particular, they have demonstrated considerable capabilities for the assessment of a wide range of tasks across multiple domains, exhibiting high levels of performances \cite{CARUCCIO2024121186,song2024llm}.

In this section, we first present the datasets used in our study and we provide an overview of the techniques and models used for augmented data.

\subsection{Datasets Overview}\label{sec:Datasets}
In this study, we adopt two datasets, i.e., \Adult, and \Italia datasets, which are some of the most used datasets for investigating the effectiveness of anonymization approaches. 
The first dataset considered in our study, namely \Adult\footnote{\label{footnoteAdult}\href{www.kaggle.com/datasets/uciml/adult-census-income}{www.kaggle.com/datasets/uciml/adult-census-income}}, also known as the \textit{Census Income Dataset}, contains $48,842$ rows collected from the $1994$ United States census database, and $15$ attributes, representing various demographic and employment-related features. The second dataset, namely \Italia, contains $100$ rows collected for demographic or epidemiological studies, and for analyzing the distribution of certain diseases across different age groups and geographic areas. It contains $4$ attributes, i.e., \textit{age}, \textit{city birth}, \textit{zip code}, and \textit{disease}. Table \ref{tab:info-datasets} shows the details of the features for \Adult and \Italia datasets. For each attribute, we provide an overview of the distributions of the instances.

\subsection{Large Language Models Overview}\label{sec:Overview_LLM}
Nowadays Large Language Models (LLMs) have demonstrated remarkable capabilities in multiple fields, such as healthcare, economics, security, and privacy. Recently, several LLMs have been released, each with its characteristics tailored to specific tasks and applications. However, in order to assess their performance, it is important to investigate their generation and comprehension capabilities for different purposes. 
To this end, we have investigated the performance of some of the latest LLMs: ChatGPT, one of the largest proprietary state-of-the-art models, and the two most recent versions of Claude 3.0, namely Sonnet and Haiku, one of the latest LLMs released in December 2023.
\noindent
\textit{ChatGPT} is an LLM developed by OpenAI
\footnote{\href{https://chat.openai.com/}{www.chat.openai.com}}
and it exhibits an expansive vocabulary and contextual awareness, enabling it to comprehend and generate human-like text across a diverse range of topics. 
\noindent
\textit{Claude 3 Sonnet and Haiku} are two new LLMs proposed by Anthropic
\footnote{\href{https://www.anthropic.com/news/claude-3-family}{www.anthropic.com/news/claude-3-family}} 
which are included in the Claude 3 model family. They demonstrate an enhanced ability to analyse, create sophisticated content, and generate code. Each model offers high levels of performance, allowing users to choose the optimum combination of intelligence, speed, and cost for their specific application.

\begin{table}[t!]
\centering
\renewcommand{\arraystretch}{1} 
\setlength{\tabcolsep}{2pt}
\scalebox{0.75}{
\begin{tabular}{llrlllrlllr}
\multicolumn{7}{c}{\large \Adult Dataset} &  & \multicolumn{3}{c}{\large \Italia Dataset} \\ \cline{1-3} \cline{5-7} \cline{9-11} 
\multicolumn{1}{c}{\textbf{Attribute}} & \multicolumn{1}{c}{\textbf{Values}} & \multicolumn{1}{c}{\textbf{\#}} & \textit{} & \multicolumn{1}{c}{\textbf{Attribute}} & \multicolumn{1}{c}{\textbf{Values}} & \multicolumn{1}{c}{\textbf{\#}} & 
\hspace{0.5cm}
& \textbf{Attribute} & \textbf{Values} & \multicolumn{1}{l}{\textbf{\#}} \\ \cline{1-3} \cline{5-7} \cline{9-11} 
\multirow{5}{*}{\textbf{\begin{tabular}[c]{@{}l@{}}Native\\ Country\end{tabular}}} & United-States & 27504 &  & \multirow{5}{*}{\textbf{Occupation}} & Professional & 8042 &  & \multirow{3}{*}{\textbf{Age}} & 1 - 50 & 55 \\
 & North America & 826 &  &  & Blue-collar & 10918 &  &  & 51 - 75 & 25 \\
 & Asia & 1039 &  &  & Administrative & 3721 &  &  & 76 - 100 & 20 \\ \cline{9-11} 
 & Europe & 520 &  &  & Sales & 3584 &  & \multirow{4}{*}{\textbf{City Birth}} & Southern Italy & 19 \\
 & Other & 273 &  &  & Service & 3999 &  &  & Northern Italy & 35 \\ \cline{1-3} \cline{5-7}
\multirow{5}{*}{\textbf{Education}} & High School or less & 13335 &  & \multirow{2}{*}{\textbf{Sex}} & Male & 20380 &  &  & Central Italy & 37 \\
 & Some college & 7686 &  &  & Female & 9782 &  &  & Islands & 9 \\ \cline{5-7} \cline{9-11} 
 & Bachelor's & 5044 &  & \multirow{4}{*}{\textbf{\begin{tabular}[c]{@{}l@{}}Marital\\ Status\end{tabular}}} & Married & 14456 &  & \multirow{4}{*}{\textbf{Zip Code}} & 0 - 25K & 21 \\
 & Advanced degree & 2544 &  &  & Never-Married & 9726 &  &  & 25K - 50K & 36 \\
 & Other & 553 &  &  & Divorced & 5153 &  &  & 50K - 75K & 13 \\ \cline{1-3}
\multirow{4}{*}{\textbf{Workclass}} & Private & 22286 &  &  & Widowed & 827 &  &  & 75K - 100K & 22 \\ \cline{5-7} \cline{9-11} 
 & Self-employed & 3573 &  & \multirow{4}{*}{\textbf{Age}} & 18-25 & 5290 &  & \multirow{6}{*}{\textbf{Disease}} & Heart disease & 21 \\
 & Government & 4289 &  &  & 26-35 & 8054 &  &  & Anorexia & 21 \\
 & Other & 14 &  &  & 36-45 & 7734 &  &  & Autism & 20 \\ \cline{1-3}
\multirow{3}{*}{\textbf{Race}} & White & 25933 &  &  & 46$+$ & 9084 &  &  & AIDS & 14 \\ \cline{5-7}
 & Black & 2817 &  & \multirow{2}{*}{\textbf{\begin{tabular}[c]{@{}l@{}}Salary\\ Class\end{tabular}}} & $\leq$50K & 22654 &  &  & Alzheimer  & 13 \\
 & Other & 1412 &  &  & $\geq$50K & 7508 &  &  & Cancer & 11 \\ \cline{1-3} \cline{5-7} \cline{9-11} 
\end{tabular}
}\vspace{0.2cm}
\caption{Overview of the features of \Adult and \Italia datasets. \label{tab:info-datasets}}
\end{table}

\subsection{Data Anonymization Tecniques}\label{sec:techniques}
Data anonymization techniques include all techniques used to protect the privacy of individuals in a dataset. They include ensuring that no individual can be identified by generalizing the data, masking the data, or replacing Personally Identifiable Information (PII) with a pseudonym. One of the most famous anonymization techniques is the k-anonymity \cite{sweeney2002k}. The principle of k-anonymity is to ensure that each individual in the dataset cannot be distinguished from at least $k-1$ other individuals, thus preserving their anonymity. This is typically achieved by generalizing or suppressing certain identifying attributes, also known as quasi-identifiers (QIs), such as age or postal code, to a level where at least $k$ records share the same values for these attributes. In particular, the QIs is defined as a set of attributes that, combined with internal/external information, can be used to re-identify all or some of the individuals to whom the information relates.
There are several k-anonymity algorithms used to perform anonymization, such as \textit{Basic Mondrian}\footnote{\label{footnote1}\href{https://github.com/fhstp/k-AnonML}{https://github.com/fhstp/k-AnonML}}, \textit{Top Down Greedy Anonymization}\footref{footnote1}, and \textit{Clustering Based k-Anonymization}\footref{footnote1}.
In particular, the Basic Mondrian \cite{1617393}, is a top-down greedy data anonymization algorithm for relational datasets that is tailored to both categorical and numerical data by using generalization hierarchies.
The Top Down Greedy Anonymization (TDGA) \cite{TopDown} is a greedy algorithm that begins by considering the entire dataset and then recursively divides it into smaller subsets. This process is guided by a greedy approach. The unique aspect of TDGA is its use of a binary splitting and balancing strategy that involves either merging subgroups or moving some records from a larger group to a smaller one. This strategy aims to balance the distribution of data between the different groups, thus increasing the usefulness of the anonymized data. In comparison to other anonymization methods such as Mondrian, TDGA has been shown to preserve greater data utility. 
Concerning the Clustering-Based k-Anonymisation (CBA)\cite{clustering}, it aims to partition the tuples of a given dataset into a large number of clusters, typically $k$ clusters in the case of k-anonymity, by using k-Nearest Neighbor (KNN) or the K-Member algorithms. This is achieved by ensuring that individuals within a cluster are more similar to each other than individuals in different clusters. 
In our study, we used these algorithms to anonymize \Adult and \Italia datasets considering different values of $k$.

\section{Anonymized Data Augmentation with LLMs} \label{sec:methodology}
In today's data-driven world, the need to protect individuals' privacy while utilizing large datasets for analysis is a fundamental task for researchers and organizations. The latter, in fact, often possess vast amounts of sensitive information that, if improperly handled, can lead to privacy breaches. 
The advent of LLMs could present a novel approach to tackling this issue. In fact, these models offer potential solutions for enhancing data anonymization processes with their ability to understand and extract complex patterns in data. 
In this section, we first provide an overview of {anonymized data augmentation} data with LLM using the k-anonymity approach, and then we discuss the strategy for defining multimodal prompts for interacting with LLMs.   

\subsection{Problem Overview}
Anonymizing data is a critical process aimed at preventing the identification of individuals from datasets by transforming the data in such a way that personal identities are concealed. Although a large number of approaches and frameworks have been proposed, the need to find efficient and reliable methods remains a widely analyzed goal, since traditional anonymization techniques often face challenges in maintaining the balance between privacy and data utility. One of the significant problems of using anonymization techniques is that they can create data that becomes less useful or even unusable for predictive models, affecting the overall value of the data for analytical purposes. This limitation underscores the necessity of developing methods to augment data while preserving anonymization, ensuring that the data remains valuable for predictive models. 

Large Language Models (LLMs) hold significant potential in achieving this goal, since they have demonstrated remarkable capabilities in generating high-quality synthetic data that can replicate real datasets \cite{yu2024large}. In fact, LLMs are able to understand the underlying patterns and distributions in the data, allowing them to generate synthetic data that retains the statistical properties and predictive power of the original data. However, it is necessary to investigate their capabilities to generate data when used to augment anonymized datasets. 
The problem of using LLMs to augment anonymized data involves the need to ensure that the augmented data maintains specific privacy criteria, such as k-anonymity, while maintaining its usefulness for analysis.
To this end, we investigate the capabilities of LLMs to augment anonymized data according to a k-anonymity approach considering different values for the $k$. 

Formally, let $\Delta$ be the initial dataset consisting of $n$ records $\Delta = \{R_1, R_2, \dots,$ $R_n\}$, where each record $R_i$ contains $m$ attributes, i.e., $R_i = \{A_{1_i}, A_{2_i}, \dots,$ $A_{m_i}\}$. Moreover, let us consider $\Delta'$ as the k-Anonymized dataset $\Delta$ with a certain value of $k$ and containing $m$ attributes. We can identify a subset of attributes of $m$ named \textit{quasi-identifiers}, i.e., $\Psi \subset \{A_1, A_2, \ldots, A_{z}\}$, containing $z$ attributes of $m$ with $z < m$. These are non-sensitive attributes that, when combined, can uniquely identify an individual. 
Let $S$ be a sensitive attribute with $S \in \{A_1, A_2, \ldots, A_{m-z}\}$, the remaining attributes are considered to be non-sensitive.
The dataset $\Delta'$ respects the k-anonymity if each record is indistinguishable from at least $k-1$ other records with respect to the quasi-identifiers. 
Starting from this, the problem of augmenting anonymized data with an LLM $\Lambda$ aims to define a new set of new records $\Delta'' = \{R''_1, R''_2, \ldots, R''_j\}$ which can be merged with $\Delta'$ without affecting its anonymity level. The merged dataset is defined as $\Delta^* = \{\Delta' \cup \Delta''\}$, and it must satisfy or improve the k-anonymity level defined in $\Delta'$. 

\subsection{Multimodal Prompt Engineering Approach}\label{sec:Multimodal} 
The interaction with LLMs for specific tasks requires the design of ad-hoc prompt engineering templates to achieve reliable results in the context of the study. To this end, we propose two different prompting template functions in order to interact with different LLMs, one to allow LLMs to understand the context of the task, and to analyze the data, and one to allow them to understand the instructions for the required task. In both prompts, we employ the Manual Template Engineering approach to generate prompts, considered the most intuitive method for crafting templates derived from human insights \cite{liu2023pre}.\\
\noindent
\textit{Context Understanding Prompt.} The first prompting function aims to support LLMs in the comprehension of the anonymized dataset with the aim to collect the quasi-identifiers, the associated taxonomies, and the sensitive attribute, given the anonymized dataset as attached file.
In fact, as recently demonstrated, by providing context to LLMs before performing any task, they are able to provide answers and perform tasks more accurately \cite{caruccio2024claude,ding2024survey}.
To this end, we have defined a new multimodal prompting function $f_{\text{context-understanding}}(x)$ that aims to provide a context to the LLM and complete the sentence $x$ considering both textual and files as input, and achieve the prompt sentence $x' = f_{\text{context-understanding}}(x)$. 
The template of this function has been defined as follows:
\begin{tcolorbox}[enhanced, colback=white,width=1\linewidth, center upper, halign=left, left=2mm, right=2mm,fontupper=\scriptsize, drop shadow southwest, sharp corners,  title={Template of $f_{\text{context-understanding}}(\Delta)$}, coltitle=black, boxed title style={opacityback=0,colframe=white,size=fbox,arc=0mm}, attach boxed title to top left={yshift=-\tcboxedtitleheight/2,xshift=3mm}]
   \centering
 Given this attached dataset [$\Delta$], I have performed a Data Anonimization task by using the K-Anonimity technique. Please provide a comprehensive description of it, including the following details:
 - Dataset Overview: Number of rows (records) and columns (attributes); Purpose and context of the data.
  Moreover, provide the following column details: \\
  List all column names and for each of them, specify the data type (numerical, categorical, etc.); Identify the quasi-identifier columns whose values could potentially re-identify individuals when combined; Specify the sensitive column containing confidential information requiring protection; Identify the taxonomy used to anonymize this dataset. [$\Gamma$], [S]   \end{tcolorbox}
\noindent
where [$\Delta$] is the slot of the attached file containing the original dataset, [S] the slot of the sensitive attribute identified by the $\Lambda$, and [$\Gamma$] is the slot containing the pairs $\{\langle A_{i}, \tau_{i} \rangle | A_{i} \in \Psi, i = 1, 2, \ldots, z\}$ with $\tau_{i}$ representing the taxonomy associated with $A_i$. The taxonomies are the hierarchical relationships and generalization levels of \(A_i\), which will be used by $\Lambda$ for generating new data.\\
\textit{Data Augmentation Prompt.} The second prompting function aims to generate the new anonymized data to merge with the original anonymised dataset, given the quasi-identifiers, their taxonomies, and the sensitive attribute. Thereby, we have defined a prompting function $x''=f_{\text{augmented-anonymized-data}}(x)$ to generate new anonymized data. The template of this function has been defined as follows:
 \vspace{-0.45cm}
\begin{tcolorbox}[enhanced, colback=white,width=1\linewidth, center upper, halign=left, left=2mm, right=2mm, fontupper=\scriptsize, drop shadow southwest, sharp corners,  title={Template of $f_{\text{augmented-anonymized-data}}(\Delta', k, S, \Gamma)$}, coltitle=black, boxed title style={opacityback=0,colframe=white,size=fbox,arc=0mm}, attach boxed title to top left={yshift=-\tcboxedtitleheight/2,xshift=3mm}]
\centering
Given the attached anonymized dataset [$\Delta'$] and the following information: 1. Quasi-identifiers and their taxonomies: [$\Gamma$]; 2. A Sensitive attribute: [S]; Current k-anonymity level: k $=$ [K].

Task: Generate new records that can be merged with the original dataset while satisfying the following conditions: a) The new records must follow the same structure and data types as the original dataset; b) The values for quasi-identifiers must be consistent with the provided taxonomies; c) The values for the sensitive attribute must be consistent with the dataset.\\

Important considerations: 1) Ensure that the new records do not introduce new unique combinations of quasi-identifiers that could reduce anonymity; 2) The distribution of sensitive attribute values in the new records should be similar to the original dataset to prevent attribute disclosure. 3) If possible, aim to increase or maintain the overall k-anonymity level of the merged dataset. $[R''_1], {...}, [R''_j]$
\end{tcolorbox}
\noindent
where slot [$\Delta'$] is the attached file containing the anonymized dataset with a value of $k$ equal to [K], which has to be augmented.
The slots $[R''_1], \dots, [R''_j]$ containing the newly generated records by $\Lambda$ that merged with $\Delta'$ enable us to obtain the new augmented dataset $\Delta^*$.
More specifically, this interaction with LLMs aims to generate a set of augmented tuples, i.e., tuples that already are in the dataset or new fake tuples, in order to maintain the k-anonymity level of the dataset. 
It is important to note that the number of tuples generated by an LLM can exhibit variability from one interaction to another. This variability probably stems from multiple factors related to the prompt and dataset, as well as the statistical nature of the models.

\section{Experimental Evaluation}\label{sec:exp}
In this Section, we outline the experimental evaluation performed in order to investigate the capability of LLMs to augment anonymized data. In particular, we first provide an overview of the experimental settings and metrics used to assess the new augmented data. Then, we analyze the capabilities of LLMs to highlight their strengths and limitations in augmenting anonymized data.

\paragraph{Experimental Settings.}
The experimental evaluations have been conducted using the datasets shown in Section \ref{sec:Datasets}. We perform different experimental sessions to assess the capabilities of LLMs in various scenarios, using specific values for k-anonymity. We conduct experiments with $k = 2$, and then with $k$ values ranging from $5$ to $100$ with an increment of $5$.
For each resulting dataset, we have applied the k-anonymity algorithms, discussed in Section \ref{sec:techniques}, by considering different values of $k$. 
To check and assess the level of anonymity of a dataset through k-anonymity, we employ the pyCanon \cite{sainzpardo2022pyCanon} library, by giving the set of quasi-identifiers and a sensitive attribute collected by the interaction with LLMs.

Concerning LLMs involved in our study, we used the official platforms for proprietary models, i.e., for which the source code has not been released or with high hardware requirements, such as ChatGPT, Claude 3.0 Sonnet, and Haiku. To enable them to process datasets, we have selected a $600$ random tuple from each dataset, since their limits in the size of prompts do not allow us to process a larger amount of data.
All the experiments have been executed on a workstation with an Intel i9 CPU at $5$ GHz, $14$-core, and $64$GB of memory, equipped with a GPU NVIDIA 3060 GPU.

\subsection{Experimental Results}
Ensuring the anonymity of a dataset remains a significant challenge, largely because each dataset is unique and tailored to address specific problems. In this context, augmenting anonymized data using generative models introduces additional complexities. This is because there is a higher risk of inadvertently generating synthetic data that could compromise anonymity by making it easier to identify or disclose sensitive information.
Thus, it is necessary to evaluate the anonymity level of each dataset involved in our study, with different values of $k$, in order to compare the actual anonymization level achieved through the k-anonymity algorithms, discussed in Section \ref{sec:techniques}.

\begin{table}[t!]
\centering
\small
\renewcommand{\arraystretch}{0.9} 
\setlength{\tabcolsep}{7pt}
\scalebox{0.90}{

\begin{tabular}{lccc}
\multicolumn{4}{c}{\large \Adult Dataset} \\
\hline
\textbf{k}
& BM & TDGA & CBA\\
\hline
2 & $=$ & $=$ & $=$ \\
5 & $=$ & $=$ & $=$ \\
10 & $=$ & $=$ & $=$ \\
15 & $=$ & $=$ & $=$ \\
20 & >(22) & $=$ & $=$ \\
25 & >(26) & $=$ & $=$ \\
30 & >(35) & $=$ & $=$ \\
35 & $=$ & $=$ & $=$ \\
40 & >(46) & $=$ & $=$ \\
45 & >(46) & $=$ & $=$ \\
50 & >(61) & $=$ & $=$ \\
55 & >(61) & $=$ & $=$ \\
60 & >(61) & $=$ & $=$ \\
65 & >(163) & $=$ & $=$ \\
70 & >(163) & >(76) & $=$ \\
75 & >(163) & $=$ & $=$ \\
80 & >(163) & $=$ & $=$ \\
85 & >(163) & $=$ & $=$ \\
90 & >(163) & >(111) & $=$ \\
95 & >(163) & >(102) & $=$ \\
100 & >(163) & $=$ & $=$ \\
\hline
\end{tabular}

\quad

\begin{tabular}{lccc}
\multicolumn{4}{c}{{\large \Italia Dataset}} \\
\hline
\textbf{k} & BM & TDGA & CBA\\
\hline
2 & $=$ & $=$ & $=$\\
5 & $=$ & $=$ & $=$ \\
10 & >(11) & >(12) & $=$ \\
15 & >(17) & >(49) & $=$ \\
20 & >(26) & >(44) & $=$ \\
25 & >(26) & >(39) & >(100) \\
30 & >(100) & >(34) & >(100) \\
35 & >(100) & >(100) & >(100) \\
40 & >(100) & >(100) & >(100) \\
45 & >(100) & >(100) & >(100) \\
50 & >(100) & >(100) & >(100) \\
55 & >(100) & >(100) & >(100) \\
60 & >(100) & >(100) & >(100) \\
65 & >(100) & >(100) & >(100) \\
70 & >(100) & >(100) & >(100) \\
75 & >(100) & >(100) & >(100) \\
80 & >(100) & >(100) & >(100) \\
85 & >(100) & >(100) & >(100) \\
90 & >(100) & >(100) & >(100) \\
95 & >(100) & >(100) & >(100) \\
100 & $=$ & $=$ & $=$ \\
\hline
\end{tabular}
}
\vspace{0.2cm}
\caption{Anonimity level of \Adult and \Italia dataset with pyCanon. \label{tab:info-datasets-pyCanon}}
\end{table}
\paragraph{Anonimize Dataset.}
Table \ref{tab:info-datasets-pyCanon} shows the anonymity level of \Adult and \Italia datasets evaluated using the pyCanon python library. 
As we can see, for the \Italia dataset the algorithm that allows to achieve the request k-anonymity is the Cluster-based Anonymity (CBA). This algorithm outperforms the other one due to the use of clustering algorithms such as k-Nearest Neighbor (KNN) or K-Member that ensure the generation of exactly $k$ clusters. This method is particularly effective in maintaining the balance between data utility and privacy by optimizing the grouping of similar records. 
Concerning the use of TDGA, it can be seen that this algorithm, compared to CBA, can also efficiently apply the request level of anonymity. This can be due to the Top-Down strategy to partition the dataset, i.e., it firstly start from the entire dataset and than iteratively splitting it into smaller groups in order to achieved the k value and to optimize privacy and utility.
Otherwise, the Basic Mordian algorithm tends to produce different values of k across all generated datasets. The main idea of it is to recursively split data into groups that are as similar as possible, leading in the case of non-homogeneous datasets, to create non-optimal groups. 
In particular, CBA consistently outperforms both TDGA and BM in achieving higher levels of k-anonymity across different values of k. This is due to its ability to create well-defined clusters through the effective use of clustering algorithms, leading to better privacy protection while retaining data utility.

For the \Italia dataset, all the algorithms have been affected by the smaller dataset size, i.e., achieving k-anonymity can be more challenging because there are fewer records to group together to form homogeneous clusters. As the first dataset, the CBA algorithm outperforms the other one due to more efficiency in well-defined groups, also in case of a lower number of instances. In contrast, TDGA and BM demonstrated lower performances, which underlines the importance of selecting appropriate anonymization techniques tailored to specific features, or suitable for the specific size of the dataset.

\begin{table}[t!]
\centering
\footnotesize
\renewcommand{\arraystretch}{1.1} 
\setlength{\tabcolsep}{1pt}
\scalebox{0.58}{
\begin{tabular}{l|ccc||ccc||ccc|||ccc||ccc||ccc}

\multicolumn{9}{c}{{\large\Adult Dataset}} & \multicolumn{9}{c}{{\large\Italia Dataset}} \\
{\large k} & \multicolumn{3}{c}{{\large BM}} & \multicolumn{3}{c}{{\large TDGA}} & \multicolumn{3}{c}{{\large CBA}}
& \multicolumn{3}{c}{{\large BM}} & \multicolumn{3}{c}{{\large TDGA}} & \multicolumn{3}{c}{{\large CBA}} \\
\hline
& \parbox[t]{2mm}{\rotatebox[origin=c]{90}{Sonnet}} & \parbox[t]{2mm}{\rotatebox[origin=c]{90}{Haiku}} & \parbox[t]{2mm}{\rotatebox[origin=c]{90}{ChatGPT}}
& \parbox[t]{2mm}{\rotatebox[origin=c]{90}{Sonnet}} & \parbox[t]{2mm}{\rotatebox[origin=c]{90}{Haiku}} & \parbox[t]{2mm}{\rotatebox[origin=c]{90}{ChatGPT}}
& \parbox[t]{2mm}{\rotatebox[origin=c]{90}{Sonnet}} & \parbox[t]{2mm}{\rotatebox[origin=c]{90}{Haiku}} & \parbox[t]{2mm}{\rotatebox[origin=c]{90}{ChatGPT}}
& \parbox[t]{2mm}{\rotatebox[origin=c]{90}{Sonnet}} & \parbox[t]{2mm}{\rotatebox[origin=c]{90}{Haiku}} & \parbox[t]{2mm}{\rotatebox[origin=c]{90}{ChatGPT}}
& \parbox[t]{2mm}{\rotatebox[origin=c]{90}{Sonnet}} & \parbox[t]{2mm}{\rotatebox[origin=c]{90}{Haiku}} & \parbox[t]{2mm}{\rotatebox[origin=c]{90}{ChatGPT}}
& \parbox[t]{2mm}{\rotatebox[origin=c]{90}{Sonnet}} & \parbox[t]{2mm}{\rotatebox[origin=c]{90}{Haiku}} & \parbox[t]{2mm}{\rotatebox[origin=c]{90}{ChatGPT}}\\
\hline
2 & $=$ & $=$ & $=$ & $=$ & $=$ & $=$ & $=$ & $=$ & $=$ & >(3) & $=$ & >(3) & $=$ & $=$ & >(3) & = & = & = \\
5 & $=$ & $=$ & >(6) & $=$ & $=$ & $=$ & $=$ & $=$ & $=$ & >(7) & >(6) & >(8) & >(6) & $=$ & >(7) & = & = & >(6) \\
10 & >(11) & $=$ & >(12) & $=$ & $=$ & >(11) & $=$ & $=$ & $=$ & >(13) & >(11) & >(14) & >(11) & $=$ & >(12) & = & = & >(11) \\
15 & >(16) & $=$ & >(17) & >(16) & $=$ & >(16) & $=$ & $=$ & $=$ & >(18) & >(16) & >(19) & >(16) & $=$ & >(17) & = & = & >(16) \\
20 & >(22) & >(21) & >(23) & >(21) & $=$ & >(22) & >(21) & $=$ & $=$ & >(24) & >(22) & >(25) & >(22) & >(21) & >(23) & = & >(22) & >(21) \\
25 & >(27) & >(26) & >(28) & >(26) & >(26) & >(27) & >(26) & $=$ & $=$ & >(29) & >(27) & >(30) & >(27) & >(26) & >(28) & = & >(25) & >(27) \\
30 & >(32) & >(31) & >(33) & >(31) & >(31) & >(32) & $=$ & >(31) & $=$ & >(35) & >(32) & >(36) & >(32) & >(31) & >(33) & >(31) & = & >(32) \\
35 & >(40) & >(38) & >(41) & >(39) & >(37) & >(41) & $=$ & $=$ & >(37) & >(40) & >(37) & >(41) & >(37) & >(36) & >(38) & >(36) & >(36) & >(37) \\
40 & >(54) & >(50) & >(56) & >(52) & >(48) & >(56) & >(45) & >(42) & >(47) & >(45) & >(42) & >(46) & >(42) & >(41) & >(43) & >(41) & >(47) & >(42) \\
45 & >(47) & >(46) & >(48) & >(50) & >(48) & >(48) & $=$ & >(46) & >(47) & >(50) & >(47) & >(51) & >(47) & >(46) & >(48) & = & >(47) & >(47) \\
50 & >(63) & >(60) & >(65) & >(53) & >(51) & >(65) & $=$ & >(52) & >(54) & >(55) & >(52) & >(56) & >(52) & >(51) & >(53) & >(51) & >(57) & >(52) \\
55 & >(67) & >(64) & >(69) & >(57) & $=$ & >(69) & $=$ & <(54) & >(58) & >(60) & >(57) & >(61) & >(57) & >(56) & >(58) & >(56) & >(58) & >(57) \\
60 & >(163) & >(150) & >(170) & >(62) & $=$ & >(170) & $=$ & >(59) & >(63) & >(65) & >(62) & >(66) & >(62) & >(61) & >(63) & = & >(60) & >(62) \\
65 & >(169) & >(155) & >(175) & >(68) & >(66) & >(175) & $=$ & >(67) & >(69) & >(70) & >(67) & >(71) & >(67) & >(66) & >(68) & = & >(67) & >(67) \\
70 & >(179) & >(165) & >(185) & >(75) & >(72) & >(185) & >(71) & >(73) & >(76) & >(75) & >(72) & >(76) & >(72) & >(71) & >(73) & >(71) & >(76) & >(72) \\
75 & >(171) & >(160) & >(180) & >(79) & >(77) & >(180) & >(76) & >(78) & >(80) & >(80) & >(77) & >(81) & >(77) & >(76) & >(78) & >(76) & >(79) & >(77) \\
80 & >(181) & >(170) & >(190) & >(83) & >(81) & >(190) & $=$ & >(83) & >(84) & >(85) & >(82) & >(86) & >(82) & >(81) & >(83) & >(81) & >(84) & >(82) \\
85 & >(182) & >(175) & >(190) & $=$ & >(87) & >(190) & $=$ & <(84) & >(88) & >(90) & >(87) & >(91) & >(87) & >(86) & >(88) & = & = & >(87) \\
90 & >(180) & >(175) & >(185) & >(120) & >(110) & >(185) & >(92) & >(95) & >(98) & >(95) & >(92) & >(96) & >(92) & >(91) & >(93) & >(91) & >(93) & >(94) \\
95 & >(184) & >(180) & >(190) & >(120) & >(115) & >(190) & >(97) & >(97) & >(103) & >(100) & >(97) & >(101) & >(97) & >(96) & >(98) & >(99) & >(97) & >(101) \\
100 & >(187) & >(183) & >(195) & >(124) & >(120) & >(105) & >(193) & >(102) & >(108) & >(119) & >(122) & >(116) & >(112) & >(110) & >(103) & >(105) & >(102) & >(102) \\
\hline
\end{tabular}
}
\vspace{0.2cm}
\caption{Anonimity level of the augmented \Adult and \Italia datasets with pyCanon. \label{tab:augmented-datasets-pyCanon}}
\end{table}

\paragraph{Augmented Anonymized Dataset.}
Table \ref{tab:augmented-datasets-pyCanon} shows the anonymity level of the augmented \Adult and \Italia datasets evaluated with pyCanon with different $k$ values through Claude 3 Sonnet, Haiku, and ChatGPT. 
As we can see, among all the algorithms, for the \Adult dataset the CBA has allowed LLMs to better generate augmented records with the requested k values. In particular, Sonnet demonstrated better capabilities to generate coherent augmented records that can achieve a requested k-value. Concerning Haiku and ChatGPT, they generate less consistent records compared to Sonnet, especially for higher values of k. 
For TDGA and BM algorithms, all three LLMs achieved higher values of $k$, especially with values of k more than $50$. This suggests that CBA provides a more suitable initial dataset for the generations of coherent new augmented records with LLMs while preserving anonymity constraints.
As we can observe, for the \Italia dataset across all three anonymization algorithms, all the LLMs have not effectively maintained the requested $k$. For BM, all the LLMs have generated, as $k$ increases, higher £ values than requested. With the use of TDGA algorithms, we can see an improvement in the performances for lower values of £, especially for Claude 3 Haiku. Sonnet and ChatGPT still tended to exceed the target k-value, but generally by smaller than their performance with BM. 
The CBA showed the best overall performance, allowing the LLMs to generate data closest to the requested k-anonymity levels. Sonnet and Haiku performed particularly well with CBA, achieving the exact k-value in multiple instances, especially for lower to mid-range $k$ values. ChatGPT also showed improved performance with CBA compared to the other algorithms but still tended to generate higher value of k-value more often than Sonnet or Haiku. {\colortext The size and complexity of the two dataset can affect the performance of the LLM in the generation task. }
With the \Adult dataset, the models were generally more successful in maintaining the desired anonymity levels, due to the higher number of instances. However, for the \Italia dataset, with lower data, all LLMs have achieved the requested k-anonymity levels across all anonymization algorithms with more difficulties.
Moreover, we can notice that as $k$ values increased, the performance of all LLMs in maintaining the request k-anonymity levels decreased. This can be due to when using the Top-Down and Basic Mondrian algorithms, where LLMs often generated records with higher $k$ values than requested, especially for $k$ values greater than 50.
In conclusion, we can notice that our approach is 

\section{Conclusion and Future Work} \label{sec:conclusion}
In this paper, we highlight the versatility and the capabilities of some of the most recent LLMs, i.e., Claude 3 Sonnet and Haiku, and ChatGPT, in addressing the challenge of augmenting anonymized datasets. Through extensive experimental sessions, we investigated the performances of these models using various values for $k$ and different k-anonymity clustering techniques, on  \Adult and \Italia datasets.
To interact with LLMs, we designed new multimodal prompting functions based on a manual prompt engineering approach, which allowed the first, to enable LLMs to comprehensively understand the context of the considered data, extracting quasi-identifiers, associated taxonomies, and sensitive attributes; and second, to generate new anonymized rows for augmentation of the original anonymized data.
Among the LLMs evaluated, Claude 3 Sonnet demonstrated high performance, particularly with CBA-anonymized data. This model exhibited a remarkable ability to generate augmented records that are too close to the requested k-anonymity levels. The effectiveness of LLMs in generating new records that can maintain the k-anonymity levels can be affected by the anonymization algorithm used and the characteristics of the dataset. Across both the Adult and Italia datasets, the Clustering-Based Anonymization algorithm consistently outperformed the other one, providing the most suitable datasets for LLMs to generate coherent and privacy-preserving augmented records.

In the future, we would like to investigate other anonymization methods with LLMs, such as differential privacy, which might also yield significant improvements in data utility and privacy. Furthermore, we want to extend the evaluation to a wider range of datasets and from different domains, which would provide a more comprehensive understanding of the LLMs' generalizability and robustness in anonymized data augmentation tasks.

\begin{credits}
\subsubsection{\ackname} This work was partially supported by project SERICS (PE00000014) under the NRRP MUR program funded by the EU - NGEU.

\end{credits}

\bibliographystyle{splncs04}
\bibliography{bibliography}

\end{document}